\documentclass[a4paper, oneside, twocolumn, notitlepage, 10pt]{extarticle_ecoc}
\usepackage{ecoc}
\usepackage{pgfplotstable}
\usepackage{graphicx}
\usepackage{afterpage}
\usepackage{upgreek}
\usepackage{multirow}
\usepackage[position=top]{subfig}
\usepackage{tikz,pgfplots,subfig, filecontents}
\usepackage{graphicx}
\usepackage{multicol}
\usepackage{colortbl}%
  \newcommand{\myrowcolour}{\rowcolor[gray]{0.925}}
\usepackage{longtable}
\newcommand{\mytab}{
\begin{tabular}{c c c c c c c c c }
\hline
Type    & \begin{tabular}[c]{@{}c@{}}Latency\\ ($\mu s$)\end{tabular} & \begin{tabular}[c]{@{}c@{}}Clock\\ Frequency\\ (MHz)\end{tabular} & BRAM & SRL & \begin{tabular}[c]{@{}c@{}}DSP \\ Slices\end{tabular} & LUT & FF  & \begin{tabular}[c]{@{}c@{}}Energy\\ $($nJ/bit$)$\end{tabular} \\ 

\hline\hline

TDCE & 0.55 & 250 & 23  & 0 & 18 & 5168 & 8924 & \textcolor{red}{1.47} \\

\myrowcolour%
FDE  & 13.77  & 250 & 8   & 0 & 14 & 2800 & 4755 & 2.78  \\ \hline
\end{tabular}
}

\begin{document}
\selectlanguage{english}    


\title{FPGA Implementation of Complex Value-based Clustering Filter for Chromatic Dispersion Compensation in Coherent Metro Links with Ultra-low Power Consumption}%


\author{
     Geraldo Gomes\textsuperscript{(1)}, Pedro Freire\textsuperscript{(1)}, Jaroslaw E. Prilepsky\textsuperscript{(1)},  Sergei K. Turitsyn\textsuperscript{(1)}
}

\maketitle                  


\begin{strip}
    \begin{author_descr}

 \textsuperscript{(1)}~Aston University, UK,
   \textcolor{blue}{\uline{freiredp@aston.ac.uk}}

    \end{author_descr}
\end{strip}

\renewcommand\footnotemark{}
\renewcommand\footnoterule{}


\begin{strip}
    \begin{ecoc_abstract}
        This paper introduces a new machine learning-assisted chromatic dispersion compensation filter, demonstrating its superior power efficiency compared to conventional FFT-based filters for metro link distances. Validations on FPGA confirmed an energy efficiency gain of up to 63.5\% compared to the standard frequency-domain chromatic dispersion equalizer. \textcopyright2024 The Author(s)
    \end{ecoc_abstract}
\end{strip}
\section{Introduction}
\vspace{-1mm}

Energy consumption is a key concern in optical transmission systems \cite{agrell2016roadmap,radovic2024power,freire2023low}. In high-end optical engines, digital signal processing (DSP) typically consumes about 70\% of the power (50\% in coherent pluggables) \cite{Ref02}. The chromatic dispersion compensation (CDC) block within DSP is a major power consumer. While CDC is well-established, reducing its complexity remains an active research area. Recently, chirp filtering has been proposed to reduce mathematical operations under certain conditions \cite{chirp-filtering}, and uniform quantization of filter taps has been explored to decrease operations \cite{uniform-quantization}. However, these studies lack hardware implementations validating energy efficiency. Other studies were implemented in hardware~\cite{finite-fields,power-eff-cdc}, but efficiency was evaluated for short fiber lengths and no standard metric was employed.

\begin{figure*}[b]
    \centering
    \hspace{-5mm}
    \includegraphics[height=6.4cm]{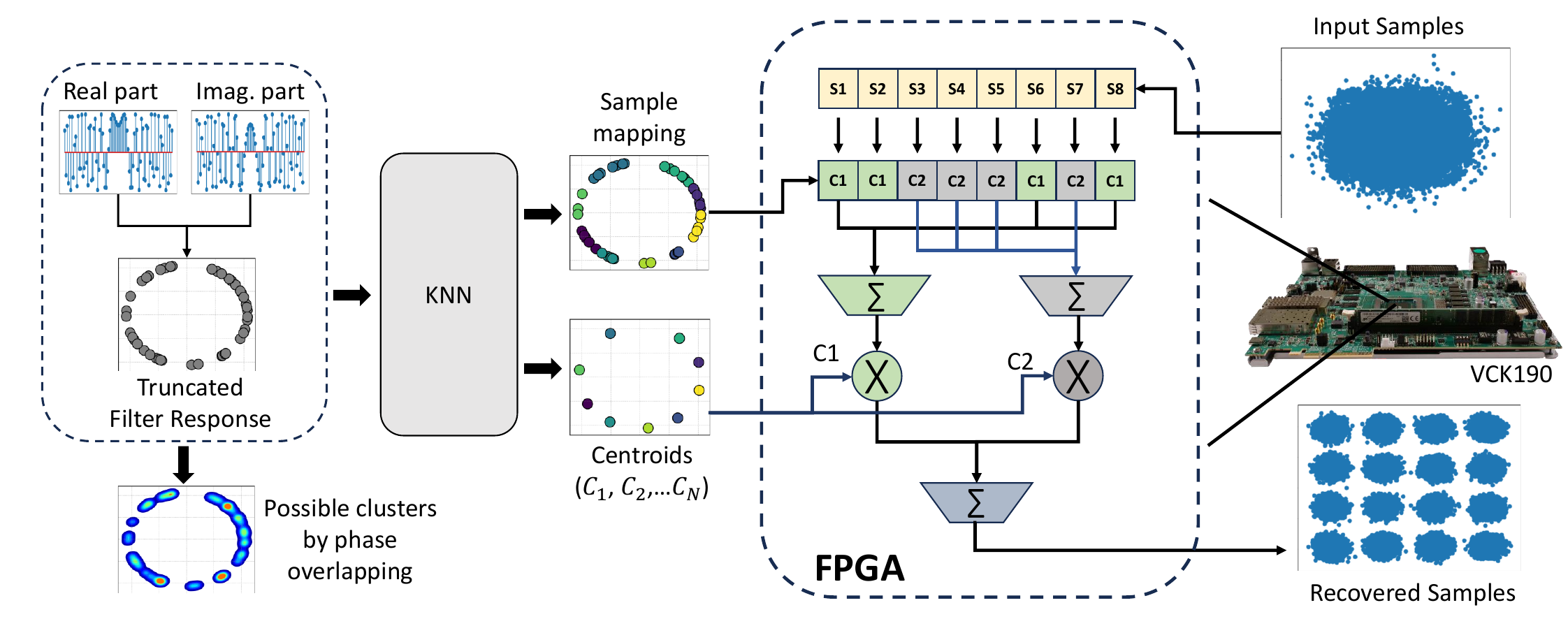}
    \caption{Proposed architecture for cluster filtering (TDCE) with simplified number of clusters for the sake of visualization}
    \label{cluster-motivation}
\end{figure*}

This paper presents a novel clustered CDC filter, which we named time domain clustered equalizer (TDCE), demonstrating its excellent power efficiency. We go beyond theoretical proposals and \textit{present hardware implementation to demonstrate the efficacy of the new CDC in decreasing power consumption across different fiber lengths.} These are nontrivial steps because additional factors in the hardware design, such as memory usage and required quantization, can affect energy consumption.  Apart from employing a clustered approach, another novel elements in our work are the analysis in terms of the \textit{nJ/recovered bit}, which is an important industrial design metric, and the fiber lengths analyzed (up to 640 km). We compare our solution with the state-of-the-art frequency domain equalizer (FDE)~\cite{txu-overlap} based on the Fast Fourier Transform (FFT) algorithm~\cite{cooley-tukey} implemented on FPGA. We also examine hardware implementation details that affect energy consumption other than the number of multiplications/summations.

\vspace{-1mm}
\section{Clustering in complex plane}
For the first time, we propose the clustering of time-domain CDC filter taps in the complex plane to reduce the CDC complexity. This approach is based on the well-known observation that the CDC filter equation for tap $g_k$
 presents a quadratic phase shift (within the context of fiber optics)\cite{savory-filter}: 
\begin{equation}
    {g}_k = \sqrt{\frac{jcT^2}{D\lambda^2 z}} \exp\left(-j \frac{\pi c T^2}{D\lambda^2 z}k^2\right)\label{taps-equation},
\end{equation}
where \(z\) is the fiber length, \(k\) is the filter tap index, \(j\) is the imaginary unit, \(D\) is the dispersion coefficient, \(c\) is the speed of light, \(\lambda\) is the wavelength of the carrier, and \(T\) is the sampling period.

As the absolute phase values increase due to the dispersion, many phases will be repeated or have values very close to each other on the unit circle in the interval between \(-\pi\) and \(+\pi\). In Fig.~\ref{cluster-motivation}, we can see the overlapping of taps and possible clusters of phase shifts. In our proposed approach, we learn such clustered phase shift values using the k-nearest neighbors (KNN) algorithm \cite{zhang2016introduction} with 300 trials to find the least complex filter.

Analysis of the clusters of phase shifts makes it possible to reduce the complexity of the time domain equalizer through the utilization of factorization of the FIR filtering process, executing first the sums of the samples associated with filter taps that are grouped, and then multiplying the sum result by the filter tap (Fig.~\ref{cluster-motivation}). The complexity of the time domain equalizer (TDE) by using FIR filtering is \(C=4N\) real multiplications per recovered symbol\cite{txu-cdc-coherent}, considering a complex multiplication is done by 4 real multiplications for a \(N\) size filter. However, for the proposed TDCE, based on a clustered filter, the complexity is \(C=4N\textsubscript{C}\), in which \(N\textsubscript{C}\) is the number of clusters utilized for the filter.

As a number of recent reports \cite{fft-hw-1,fft-hw-2,fft-hw-3} on the ASIC implementations of DSP in coherent receivers utilize the FDE based on FFT to perform CDC, we will use this equalizer as a reference. We implemented the overlap-save method\cite{txu-overlap} to break the input signal into blocks before the FDE. In this context, the FDE complexity in real multiplications per recovered symbol is given by\cite{fft-optimum}: 
\begin{equation}
\label{fft-complexity}
C\textsubscript{FFT} = N \, \frac{8 \beta \log_2(N)+4}{N-M+1},
\end{equation}
in which \(N\) is the FFT size (block size), \(M\) is the filter size and \(\beta\) is a constant equal to \(1/2\) if the Radix-2 architecture is used and the size of FFT is a power of 2, or \(\beta = 3/8\) if Radix-4 architecture is used and the length of FFT is a power of 4\cite{fft-optimum}.

\section{Simulation setup}
To test the performance of the filters, we used data of numerical modeling of a single channel, 32~GBaud, 16-QAM dual-polarisation transmission at the optimum launch power over standard single-mode fiber (SSMF). The propagation was simulated using the Manakov equation and the split-step Fourier method \cite{agrawal2000nonlinear}.  The SSMF characteristics are: dispersion $D$ = 16.8 ps/(nm$\cdot$km), nonlinearity coefficient $\gamma$ = 1.2 (W$\cdot$ km)$^{-1}$, and attenuation $\alpha$ = 0.21 dB/km. Erbium-doped fiber amplifiers (with the noise figure of 4.5 dB) after each fiber span of 80 km were considered. The number of spans varied from 1 to 8.  In our study, we employed the overlap-save method for block division of the input signal for the FDE, and all filters operated at 2 samples per symbol.

To do the FPGA implementation, we used high-level synthesis (HLS) to describe the hardware in Vitis HLS 2023.2 software from AMD that allowed us to simulate the performance by reading the input data from files, simulate the FPGA equalization using the fixed point with the proposed architecture, and save the output data to assess the BER.
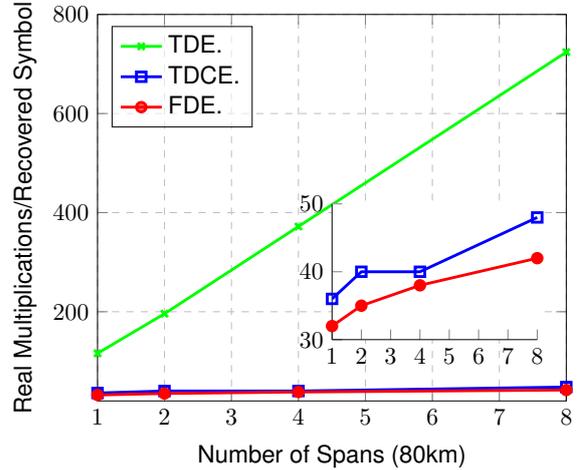
\begin{figure}[h]
    \centering
             \begin{tikzpicture}[scale=0.9]
    \begin{axis} [
        xlabel={Number of Spans (80km)},
        ylabel={Real Multiplications/Recovered Symbol},
        grid=both,  
         ylabel near ticks,
        xmin=1, xmax=8,
    	xtick={0, ..., 8},
    	ymin=20, ymax=800,
        legend style={legend pos=north west, legend cell align=left,fill=white, fill opacity=0.6, draw opacity=1,text opacity=1},
    	grid style={dashed}]
        ]
        \addplot[color=green, mark=x, very thick]     coordinates {
    (1, 116)(2, 196)(4, 372)(8, 724) 
    };
    \addlegendentry{TDE.};

            \addplot[color=blue, mark=square, very thick]     coordinates {
    (1, 36)(2, 40)(4, 40)(8, 48) 
    };
    \addlegendentry{TDCE.};
    
    \addplot[color=red, mark=*, very thick]   
    coordinates {
    (1, 32)(2, 35)(4, 38)(8, 42) 
    };
    \addlegendentry{FDE.};
        \end{axis}
       \begin{axis}[
            at={(800,290.3)},  
            anchor=north west,
            width=3cm,        
            height=2cm,       
            scale only axis,
            xmin=1, xmax=8,   
            ymin=30, ymax=50,
            xtick={1,2,3,4,5,6,7,8},  
            ytick={30,40,50}, 
            axis background/.style={fill=white},
            axis x line*=bottom,
            axis y line*=left
        ]
               \addplot[color=blue, mark=square, very thick]     coordinates {
    (1, 36)(2, 40)(4, 40)(8, 48) 
    };
   
    \addplot[color=red, mark=*, very thick]   
    coordinates {
    (1, 32)(2, 35)(4, 38)(8, 42) 
    };
        \end{axis}
    \end{tikzpicture}
    \caption{Comparison between the complexities of the TDE, TDCE and FDE equalizers across the spans with a zoom of TDCE and FDE.}
    \label{n_clusters_plot}
\end{figure}

\section{Filter specifications for implementation}

The maximum filter size for the best performance is given by\cite{savory-filter} $N = 2 \, \text{round} \{ |D|\lambda^2 z/2cT^2\} +1$.


We aim for the final filter to have a $BER<3.8\times10^{-3}$, which is the error-free pre-FEC with 7\% overhead threshold\cite{agrell2018information}. However, the clustering process is an approximation that decreases the overall performance. Therefore, the performance of the truncated filter response, Eq.~(\ref{taps-equation}), was evaluated for different filter sizes, starting from the maximum allowed size and decreasing it, to identify the minimum size that satisfies BER below $1\times10^{-3}$, which is better than $3.8\times10^{-3}$, but will be cut down by the clustering process. Then, the performance for different quantities of clusters was analyzed to find the minimum number of clusters that satisfies $BER<3.8\times10^{-3}$.

The required number of clusters for 1, 2, 4, and 8 spans were 9, 10, 10, and 12, respectively. These values are much smaller than the original number of filter taps for maximum performance:  45, 89, 177, and 353, respectively.

\begin{figure*}[ht!]
\vspace{-6mm}

    \begin{minipage}[c]{\linewidth}
    
    \hspace{-6mm}
    \vspace{-11mm}
    \begin{picture}(100,20)
        \put(-5,-60){
         \begin{tikzpicture}[scale=0.55]
            \begin{axis} [
                xlabel={Number of Spans (80km)},
                ylabel={Energy (nJ)/recovered bit},
                grid=both,  
                 ylabel near ticks,
                xmin=1, xmax=8,
            	xtick={0, ..., 8},
            	ymin=0, ymax=4,
                legend style={legend pos=south east, legend cell align=left,fill=white, fill opacity=0.6, draw opacity=1,text opacity=1},
            	grid style={dashed}]
                ]
                \addplot[color=blue, mark=square, very thick]     coordinates {
            (1, 0.85)(2, 1.00)(4, 1.47)(8, 2.45) 
            };
            \addlegendentry{TDCE.};
            
            \addplot[color=red, mark=*, very thick]   
            coordinates {
            (1, 2.34)(2, 2.55)(4, 2.78)(8, 3.15) 
            };
            \addlegendentry{FDE.};
                \end{axis}
                \node[text width=3cm] at (3.1,5.1) 
            {\textcolor{red}{\textbf{(a)}}};
            \draw[red!80!pink,dashed] (0.0,1.2) -- (1,1.2);
            \draw[red!80!pink,dashed] (0.0,3.4) -- (1,3.4);
            \draw[thick, <->,red] (0.5,1.2) -- +(0.01,2.2);
                \node[text width=1cm] at (1.85,2.6) 
            {\textcolor{red}{\footnotesize $\approx$ 63\%}};
        \end{tikzpicture}    
        }
    \end{picture}
    
    \begin{picture}(110,100)
    \hspace{-1mm}
\put(110,00){ \includegraphics[width=0.75\linewidth, height=0.27\linewidth]{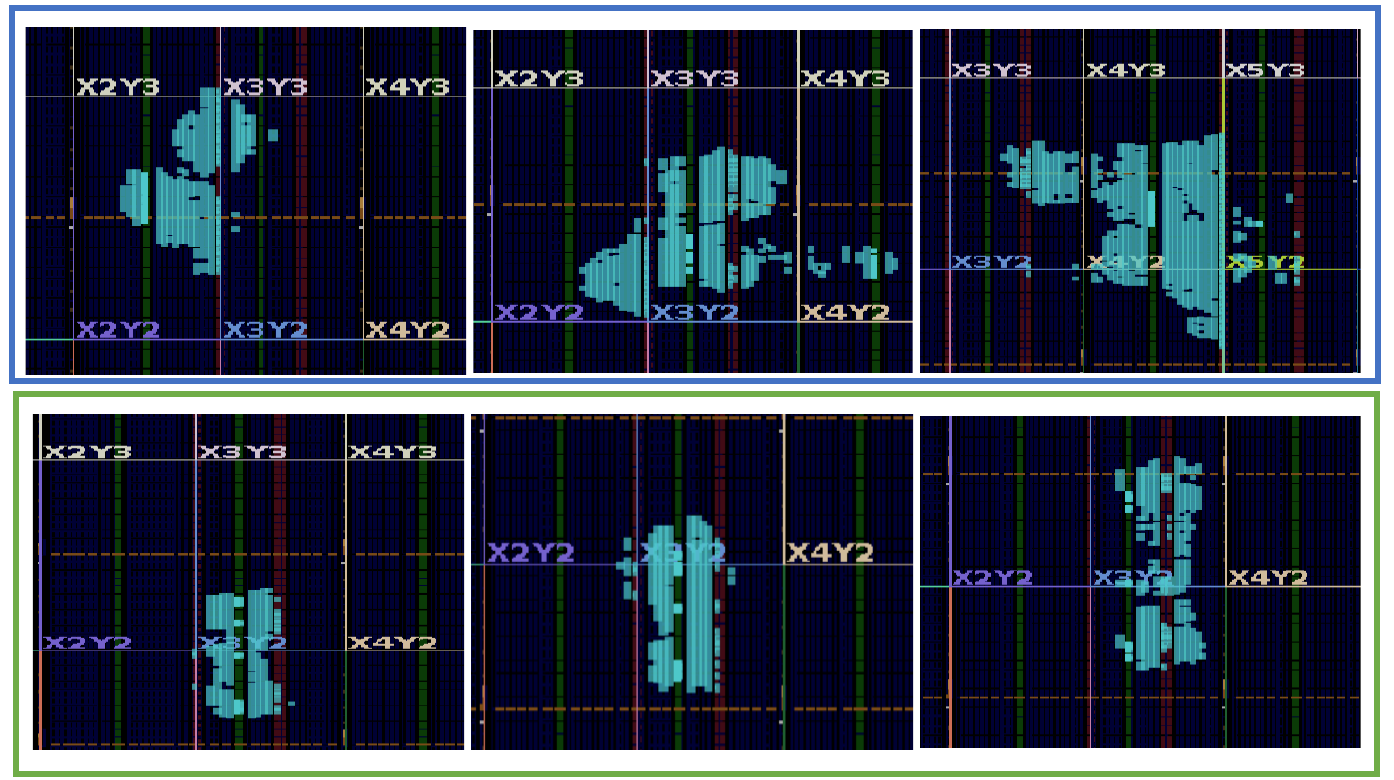}}
\put(115,125){\textcolor{red}{\textbf{(c)}}}
\put(156,125){\textcolor{red}{\textbf{2 spans}}}
\put(266,125){\textcolor{red}{\textbf{4 spans}}}
\put(376,125){\textcolor{red}{\textbf{8 spans}}}
\put(453,108){\textcolor{red}{\rotatebox{270}{\textbf{TDCE}}}}
\put(453,45){\textcolor{red}{\rotatebox{270}{\textbf{FDE}}}}
\end{picture}

    \end{minipage}

    \begin{minipage}[c]{\linewidth}
    \hspace{-7mm}
    \begin{picture}(50,50)
    \put(-2,-45){
         \begin{tikzpicture}[scale=0.55]
               \begin{axis}[
                ybar, 
                symbolic x coords={Clocks, Signals, Logic, BRAM, DSP}, 
                xtick=data, 
                ylabel={Power (mW)}, 
                ylabel style={yshift=-15pt}, 
                bar width=10pt, 
                xtick align=inside, 
                enlarge x limits=0.15, 
                legend style={at={(0.5,0.95)}, anchor=north,legend columns=-1}, 
                nodes near coords, 
                nodes near coords align={vertical}, 
                ymin=0, 
                ymax=80, 
            ]
            \addplot coordinates {(Clocks, 46) (Signals, 16) (Logic, 11) (BRAM, 19) (DSP, 33)};
            \addlegendentry{TDCE}
            \addplot coordinates {(Clocks, 42) (Signals, 56) (Logic, 44) (BRAM, 27) (DSP, 57)};
            \addlegendentry{FDE}
            \end{axis}
                \node[text width=3cm] at (3.1,5.1) 
            {\textcolor{red}{\textbf{(b)}}};
    \end{tikzpicture}
    }
    \end{picture}
    
  \begin{picture}(50,50)
\put(116,00){ \raisebox{3.1cm}{\subfloat{
 \resizebox{.695\textwidth}{.08\textwidth}{ \mytab}}}}
\put(125,78){\textcolor{red}{\textbf{(d)}}}
\end{picture}
    \end{minipage}
    \caption{TDCE versus FDE FPGA design investigation for (a) Energy [nJ/recovered bit] across spans; (b) Design power distribution for the 4 spans case. (c) Chip area utilized across different distances; (d) Resources usage for the 4 spans case.}
    \label{fig:results}
\end{figure*}
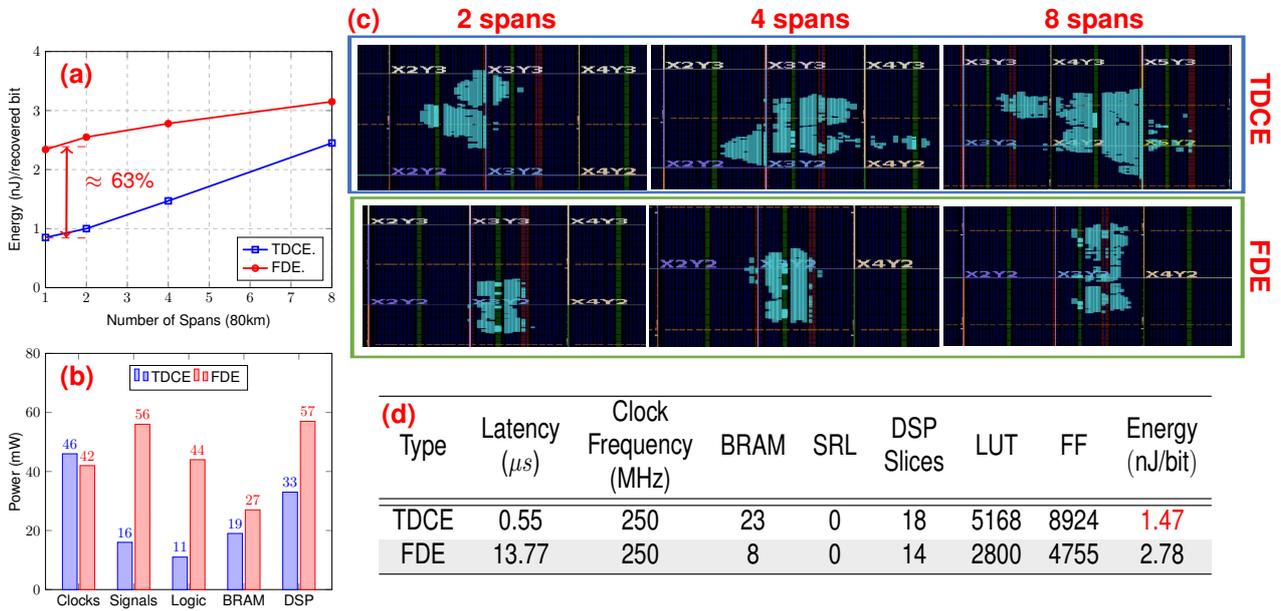

To make a fair comparison, we want to compare our solution with the lowest complexity FDE. To do so, we first found the minimum filter size that satisfies the condition $BER<3.8\times10^{-3}$, then we found the FFT size (N) that minimizes the FDE complexity by finding the minimum value of the Eq.~\ref{fft-complexity}. The optimum FFT sizes for 1, 2, 4, and 8 spans are 256, 256, 1024, and 4096, respectively for Radix-4 architecture.  

With the number of clusters and optimal FFT sizes known, we can compute the complexity depicted in Fig.~\ref{n_clusters_plot} for the TDE, FDE, and TDCE. Despite employing a time-domain approach, it is seen that we can achieve a complexity level similar to the FDE in terms of real multiplications per recovered symbol, which is significantly lower than the TDE. Next, we will move into the implementation of the two rival solutions: FDE and TDCE.

\section{FPGA implementation}
 For simplicity, we assumed that each filter is a unitary block that can be replicated to achieve higher baud rates.  All the following analyses were performed on these unitary blocks implemented on VCK190 FPGA simulated in the AMD Vitis IDE using a clock frequency of 250 MHz.
 

The FDE utilized the \textit{FFT HLS Library}\cite{XilinxFFT} developed and optimized by AMD to perform the FFT/IFFT blocks. A frequency domain multiplication block using stored filter taps was utilized after FFT and before IFFT, and then we placed a last block to remove corrupted samples due to circular convolution.


For a fair comparison, both filters utilized identical quantization for taps and signals. The minimum quantization needed to meet the pre-FEC threshold was determined for both. It was found that a total of 14 bits, with 5 for integer bits for the TDCE, and a total of 16 bits, with 1 for integer bits, were sufficient for the FDE to meet the threshold. Additionally, there's a known trade-off between digital filter throughput and hardware resources. Hence, both filters were designed to achieve the same throughput of 0.164 $\pm$ 0.012 samples per clock cycle, in all implementations across different fiber lengths to ensure fair comparison.

 To simulate the energy consumption, we synthesized the HLS code in Vitis HLS 2023.2, generated the RTL, implemented it in Vivado 2023.2 using the same default routing strategy, and assessed the energy consumption using the Power Estimator built-in tool in Vivado.  As for FPGA implementation, the static power is related to all the inner circuits regardless of whether they are in use or not; here, we only consider the dynamic power of the filters. Figure~\ref{fig:results}(a) illustrates that our approach yielded energy savings ranging from 63.5\% (1 span) to 22.4\% (8 spans) compared to FDE. Notably, for 4 spans, TDCE consumed 1.47 nJ/bit, whereas FDE consumed 2.78 nJ/bit.
 
 To gain deeper insights into our energy-saving mechanisms, we provide a more detailed breakdown of power consumption in Fig.~\ref{fig:results}(b). This analysis reveals reduced power consumption across various components such as logic, memory, DSP slices, in addition to signals and clocks.

Note that, as seen in Fig.~\ref{fig:results}(c,d), the TDCE solution utilizes more resources and occupies a larger chip area compared to the FDE implementation. Further analysis of the synthesized design revealed that despite employing more resources, TDCE switches them less frequently than FDE and utilizes simpler components making this solution more power efficient. For example, TDCE requires BRAMs with an 18-bit \textit{Write Width Port} and a 6\% \textit{Toggle Rate} \cite{toggle}, whereas FDE utilizes a 36-bit port and operates at a 38.7\% \textit{Toggle Rate}, resulting in reduced power consumption for TDCE. 

Hence,  while our solution exhibits only slightly higher complexity in terms of multiplications, it provides \textit{substantially lower energy consumption} due to factors beyond just multiplier count, such as memory organization and signal rates in the chip design.
\section{Conclusions}
We presented a novel algorithm for the CDC, implemented it on FPGA and demonstrated that even though its complexity is only slightly larger than that for the FDE, the proposed approach can provide up to 63.5\% energy saving, by analyzing its efficiency in different fiber lengths (up to 640km) using the \textit{nJ/recovered bit} metric.

\section{Acknowledgements}
PF acknowledges AMD for providing the VCK-190 board, Vivado License, and consultancy during the design of the algorithms. GG is thankful to Dr. Mike Anderson for his advice regarding power estimation. SKT acknowledges the EPSRC project TRANSNET (EP/R035342/1).


\defbibnote{myprenote}{}

\printbibliography[prenote=myprenote]

\vspace{-4mm}

\end{document}